\documentclass[12pt]{article}

\usepackage{graphicx, natbib, amsmath, amssymb, verbatim}
\usepackage{a4wide}
\usepackage{setspace}

\begin{document}
\pagenumbering{arabic}

\title{Understanding forest dynamics and plantation transformation using a simple size-structured model} 

\author{Tom Adams*$^{1}$, Graeme Ackland$^{1}$, Glenn Marion$^{2}$ and Colin
   Edwards$^{3}$} 

\date{}
\maketitle

(1) School of Physics and Astronomy, The University
  of Edinburgh, EH9 3JZ, Scotland, (2) Biomathematics and Statistics
  Scotland, EH9 3JZ, (3) Forest Research, Northern Research Station,
  Midlothian, EH25 9SY\\
* E-mail: t.p.adams@sms.ed.ac.uk




\begin{abstract}
  1. Concerns about biodiversity and the long-term sustainability of
  forest ecosystems have led to changing attitudes with respect to
  plantations. These artificial communities are ubiquitous, yet
  provide reduced habitat value in comparison with their naturally
  established counterparts, key factors being high density,
  homogeneous spatial structure, and their even-sized/aged nature.
  However, \emph{transformation} management (manipulation of
  plantations to produce stands with a structure more reminiscent of
  natural ones) produces a much more complicated (and less well
  understood) inhomogeneous structure, and as such represents a major
  challenge for forest managers.

  2. We use a stochastic model which simulates birth, growth and death
  processes for spatially distributed trees.  Each tree's growth and
  mortality is determined by a competition measure which captures the
  effects of neighbours. The model is designed to be generic, but for
  experimental comparison here we parameterise it using data from
  Caledonian Scots Pine stands, before moving on to simulate
  silvicultural (forest management) strategies aimed at speeding
  transformation.

  3. The dynamics of simulated populations, starting from a plantation
  lattice configuration, mirror those of the well-established
  qualitative description of natural stand behaviour conceived by
  \citet{oliver96}, an analogy which assists understanding the
  transition from artificial to old-growth structure.

  4. Data analysis and model comparison demonstrates the existence of
  local scale heterogeneity of growth characteristics between the
  trees composing the considered forest stands.

  5. The model is applied in order to understand how management
  strategies can be adjusted to speed the process of transformation.
  These results are robust to observed growth heterogeneity.

  6. We take a novel approach in applying a simple and generic
  simulation of a spatial birth-death-growth process to understanding
  the long run dynamics of a forest community as it moves from a
  plantation to a naturally regenerating steady state. We then
  consider specific silviculture targeting acceleration of this
  transition to ``old-growth''. However, the model also provides a
  simple and robust framework for the comparison of more general
  sivicultural procedures and goals.
\end{abstract}

\section{Introduction}

Forest stand development has been studied for many decades, and a
practical understanding of the general patterns and forms observed in
the population dynamics is well established \citep{oliver96}. However,
despite the development of a great body of simulation models for
multi-species communities \citep[e.g.][]{botkin72,pacala96,busing02},
the elucidation of general rules for the structural development of
monocultures is not clear. This is due in part to the huge variation
in physiological and morphological traits of tree species, but also
because of the importance of space and size dependent interactions.

Great progress has been made in the analysis of both size-structured
\citep[see e.g.][]{sinko67} and, more recently, spatially-structured
population models \citep[see e.g.][]{bolker97,law03}. However, an
understanding of the dynamics of real communities, structured in both
size and space, has been limited by a lack of application of simple
models, amenable to analysis and approximation, to the communities in
question \citep{gratzer04}.

An important concept in forest conservation and uneven-aged stand
management is that of ``old-growth''. This is an autogenic state which
is obtained through an extended period of growth, mortality and
regeneration, in the absence of external disturbances. It is often
seen as an ``equilibrium'' state, and is characterised by a fully
represented (high variance) age and size structure, and non-regular
spatial pattern. Depending on the species involved, it may take
several centuries to attain \citep{oliver96}. The habitat created in
this state is generally considered a paradigm of what conservation
oriented forest management might hope to achieve.

Whilst marked point process simulations have recently been used to
analyse the effects of plantation stand management
\citep{comas05,renshaw08}, we seek to develop and directly apply a
generic process-based model, which is closely related to those of
\citealp{bolker97} and \citealp{law03}), to understanding the key
elements of observed stand behaviour, from planting through to
old-growth, which can also be applied to guide silviculture. Our
approach is illustrated via application to data on Scots Pine
(\emph{L. Pinus Syslvestris}).

Transformation management aims to speed the transition to the
old-growth state, from the starting point of a plantation stand.
\citet{schutz01,schutz02} suggested methods for the attainment of this
``sustainable irregular condition''; some transformation experiments
have taken place or are in progress \citep{edwards04,loewenstein05},
whilst other work has made more in-depth analysis of the structural
characteristics of natural forest stands \citep{stoll94,mason07}. An
example of a ``semi-natural'' stand, of the type studied by
\citet{mason07} is shown in Figure 1. However,
the management history of such stands is generally not known
sufficiently (if at all) before around 100 years ago, complicating
parameter estimation and model validation.

A generic spatial, size-structured, individual based model of
interacting sessile individuals is presented in Section
\ref{sec_model}. Parameters are estimated and the model assessed using
data obtained from Scots Pine (\emph{L. Pinus Sylvestris})
communities. Section \ref{sec_results} studies model dynamics: an
initial growth dominated period gives way to a reduction in density
and a meta-stable state governed by reproduction and mortality, all of
which correspond with field observations of the growth of stands of a
range of species.  Keeping in mind this long-term behaviour, Section
\ref{sec_manage} considers examples of the application of management
practices which may accelerate transformation.

\section{Materials and Methods}
\label{sec_model}
\subsection{Model}

The model is a Markovian stochastic birth-death-growth process in
continuous (two-dimensional) space. Individuals have fixed location,
and a size which increases monotonically; these jointly define the
state space of the process.  The model operates in continuous time by
means of the Gillespie algorithm \citep{cox65,gillespie77}; this
generates a series of events (i.e.  growths, births, deaths) and
inter-event times. After any given event, the rate ($r_{event}$) of
every possible event that could occur next is computed. The time to
the next event is drawn from an exponential distribution with rate
$R=\sum r_{event}$; the probability of a particular event occurring is
$r_{event}/R$.

\subsubsection*{Interaction}

Interaction between individuals plays a key role, operating on all
population dynamic processes in the model.

Individuals interact with their neighbours by means of a predefined
``kernel'' which takes a value dependent upon their separation and
size difference. Assuming that interactions act additively, and that
the effects of size difference and separation are independent, we
define a measure of the competition felt by tree $i$
\begin{equation}
  \Phi_{i}(t) = \sum_{j \in \omega_{i}} f(s_{i}(t),s_{j}(t))g(\vec{x_{i}},\vec{x_{j}})
\end{equation}
where $\omega_{i}$ is the set of all individuals excluding $i$.
$s_{i}$ is the size of tree $i$ and $\vec{x_{i}}$ its position.

We here consider a generic form for the interaction kernel; a flexible
framework implemented by \citet{raghib06,schneider06}. Competitive
inhibition is a Gaussian function of distance to neighbours. This is
then multiplied by the size of the competitor, and a $\tanh$ function,
which represents size asymmetry in the effects of competition. That is
\begin{align}
  f(s_{i}(t),s_{j}(t))&=s_{j}(t)\left(\tanh\left(k_{s}\ln\left(\frac{s_{i}(t)}{s_{j}(t)}\right)\right)+1\right)\\
  g(\vec{x_{i}},\vec{x_{j}})&=\exp(-k_{d}|\vec{x_{i}}-\vec{x_{j}}|^{2}) \nonumber
\end{align}
where $k_{d},k_{s} \in \left[0,\infty\right)$. The $\tanh$ function
allows anything from symmetric ($k_{s}=0$) to completely asymmetric
competition ($k_{s}\to\infty$) \citep{schneider06}. Multiplying
interaction by the size of the neighbour considered reflects the
increased competition from larger individuals, independent of the size
difference (consider two tiny individuals with given
separation/size-difference, compared to two large ones with the same
separation/difference).

\subsubsection*{Growth}

We consider trees with a single size measure, ``dbh'' (diameter at
breast height (1.3m)), a widely used metric in forestry, due to its
ease of measurement in the field. Dbh has been shown to map linearly
to exposed crown foliage diameter (which governs light acquisition and
seed production) with minimal parameter variation across many species
(Purves, unpublished data, and see \citealp{larocque02}).

We use the Gompertz model for individual growth \citep{schneider06},
reduced by neighbourhood interactions \citep{wensel87}. This leads to
an asymptotic maximum size, and was found to be the best fitting,
biologically accurate, descriptor of growth in statistical analysis of
tree growth increment data (results not shown).

Trees grow by fixed increments $ds=0.001$m at a rate
\begin{equation}
  G_{i}(t) = \frac{1}{ds} s_{i}(t)\left( \alpha - \beta \ln(s_{i}(t)) -
    \gamma \Phi_{i}(t) \right)
  \label{eqn_gomp}
\end{equation}
In the absence of competition ($\Phi_{i}=0$), the asymptotic size of
an individual is thus $s^{*} = \exp(\alpha / \beta)$. Under intense
competition, the right hand side of Equation \ref{eqn_gomp} may be
negative. In this case, we fix $G_{i}(t)=0$ (similarly to e.g.
\citealp{weiner01}). Variation in $ds$ has minimal effect on dynamics
provided it is sufficiently small that growth events happen frequently
compared to mortality and birth.

\subsubsection*{Mortality}

Mortality of an established individual occurs at a rate
\begin{equation}
  \label{mort1}
  M_{i}(t) = \mu_{1} + \mu_{2}\Phi_{i}(t)
\end{equation}
$\mu_{1}$ is a fixed baseline \citep{wunder06}, and $\mu_{2}$ causes
individuals under intense competition to have an elevated mortality rate
\citep{taylor07}.

\subsubsection*{Reproduction}
\label{sec_reproduction}

Exisiting individuals produce offspring of size $s=0.01$m at a rate
determined by their seed production. This is proportional to crown
foliage area, and hence also to basal area. The individual rate of
reproduction is thus $f_{i}(t) = f \pi s_{i}(t)^{2}/4$.

Offspring are placed at a randomly selected location within 10m of the
parent tree with probability of establishment/survival
$\mathbb{P}_{e}=\left(1-(\mu_{1} + \mu_{2}\Phi_{offspring}(t))\right)^{y}$.
This approximation assumes $y$ years taken to reach initial size
(0.01m dbh) and avoids introduction of time-lagged calculations, which
would impair computational and mathematical tractability.

The fecundity of trees and accurate quantification of seed
establishment success is a long standing problem in forest ecology,
due the combination of seed production, dispersal, neighbourhood and
environmental effects involved \citep{clark04,gratzer04}. Submodels
for regeneration are often used, but due to data collection issues,
precise definition of their structure and parameterisation is more
difficult \citep[e.g.][]{pacala96}. The approximation described above
effectively removes this stage of the life cycle from the model,
allowing a focus on structure in mature individuals only.

Our presented simulations use an establishment time ($y$) of 20 years,
which is supported by field studies of Scots Pine regeneration (Sarah
Turner, unpublished data).

\subsection{Statistics}
Community structure is tracked via various metrics: density (number of
individuals per m$^{2}$), total basal area ($\sum_{i}\pi
s_{i}^{2}/4$), size and age density distributions, and pair
correlation and mark correlation functions (relative density and size
multiple of pairs at given separation, \citealp{penttinen92,law09}).
All presented model results; means and standard deviations (in
Figures, lines within grey envelopes) are computed from 10 repeat
simulation runs.

The simulation arena represents a 1ha plot ($100 \times 100$m).
Periodic boundary conditions are used. Results are not significantly
altered by increasing arena size, but a smaller arena reduces the
number of individuals to a level at which some statistics cannot be
computed accurately.

\subsection{Parameterisation}
\label{sec_paramsec}

We use data from two broad stand types (collected in Scotland by
Forest Research, UK Forestry Commission): plantation and
``semi-natural'' \citep[see][]{edwards06b,mason07}.

Plantation datasets ($6 \times 1.0$ha stands) from Glenmore (Highland,
Scotland) incorporate location and size, allowing comparison of basic
statistics at a single point in time (stand age $\approx$ 80 years).

Semi-natural data is available from several sources.  Spatial point
pattern and increment core data (measurements of annual diameter
growth over the lifespan of each tree, at 1.0m height) for four
$0.8ha$ stands in the Black Wood of Rannoch (Perth and Kinross,
Scotland) allows estimation of growth (and growth interaction)
parameters. Location and size measurements (at one point in time) from
a $1.0ha$ semi-natural stand in Glen Affric (Highland, Scotland)
provide another basis for later comparison.

In none of the stands is there adequate information to reliably
estimate mortality ($\mu_{1}$, $\mu_{2}$) or fecundity ($f$). These
are thus tuned to satisfactorily meet plantation and steady state
(semi-natural stand) density. The baseline mortality rate used gives
an expected lifespan of 250 years \citep{tfl98,forestry09}.

A nonlinear mixed effects (NLME) approach \citep{lindstrom90} was used
to estimate growth parameters $\alpha$, $\beta$ and $\gamma$.
Best-fitting growth curves were computed for each of a subset of
individuals from two of the Rannoch plots, and the mean, standard
deviation and correlation between each parameter within the population
was estimated. Details are given in Appendix 1, Electronic
Supplementary Materials (ESM). Mean values for $\alpha$ and $\beta$
are used for simulation, though large variation between individuals
was observed. $\gamma$ was difficult to estimate from the semi-natural
data, its standard deviation being larger than its mean.  However, it
has a large effect on the simulated ``plantation'' size distribution,
whilst semi-natural stand characteristics are relatively insensitive
to its precise value (Appendix 2, ESM).  Therefore a value slightly
lower than the estimated mean was used in order to better match the
size distribution in both plantation and semi-natural stages.

$k_{d}$ was selected to provide an interaction neighbourhood similar
to previous authors \citep[e.g.][]{canham04}. $k_{s}$ determines early
(plantation) size distribution, and was selected accordingly (it has
minimal effect on long-run behaviour).

All parameter values used for simulation are shown in Table
\ref{tab_parametertable}. Sensitivity to parameter variation over
broad intervals was also tested, a brief summary of which is provided
in Appendix 2, ESM.

A standard planting regime implemented in Scots Pine plantations is a
2m square lattice, typically on previously planted ground. Old stumps
and furrows prevent a perfectly regular structure being created, so
our initial condition has 0.01m dbh trees with small random deviations
from exact lattice sites.

\begin{table}[h]
  \centering
  \caption{Model parameters, description and values.}
  \vspace{0.1in}
  \begin{tabular}{l|l|l}
    Parameter & Description & Value\\
    \hline
    {\bf population rates}&&\\
    $f$ & reproductive rate per m$^{2}$ basal area & 0.2 \\
    $\mu_{1}$ & baseline mortality & 0.004 \\
    $\mu_{2}$ & mortality interaction & 0.00002 \\
    $\alpha$ & gompertz a & 0.1308\\
    $\beta$ & gompertz b & 0.03158 \\
    $\gamma$ & growth interaction & 0.00005 \\
    {\bf interaction kernels}&&\\
    $k_{d}$ & distance decay & 0.1 \\
    $k_{s}$ & size asymmetry & 1.2 \\
  \end{tabular}
  \vspace{0.1in}
  \label{tab_parametertable}
\end{table}

\section{Results}
\subsection{Model Behaviour and comparison with data}
\label{sec_results}
Starting from the plantation configuration, the model community
displays three distinct stages:
\begin{itemize}
\item {initial growth dominated period, during which the plantation
    structure largely remains}
\item {a period of high mortality and basal area reduction as the
    impact of interactions begin to be felt, together with an increase
    in regeneration as the canopy opens}
\item {the long-run meta-stable state, during which stand structure is
    more irregular and determined by the levels of mortality and birth}
\end{itemize}
Oliver and Larson's (1996) qualitative description of the development
of natural forest stands from bare ground is now well established
\citep{peterken96,wulder06}. It provides the following
characterisation of the overall behaviour of the community: stand
initiation $\rightarrow$ stem exclusion $\rightarrow$ understory
reinitiation $\rightarrow$ old growth. This is similar to our
plantation initiated model, except we find that stem-exclusion and
regeneration occur simultaneously.

The characteristics of each stage will now be discussed in more
detail. A summary of the effects of parametric variation upon key
properties is given in Appendix 2, ESM.

\begin{figure}
  \centering
  \scalebox{0.20}[0.20]{\includegraphics{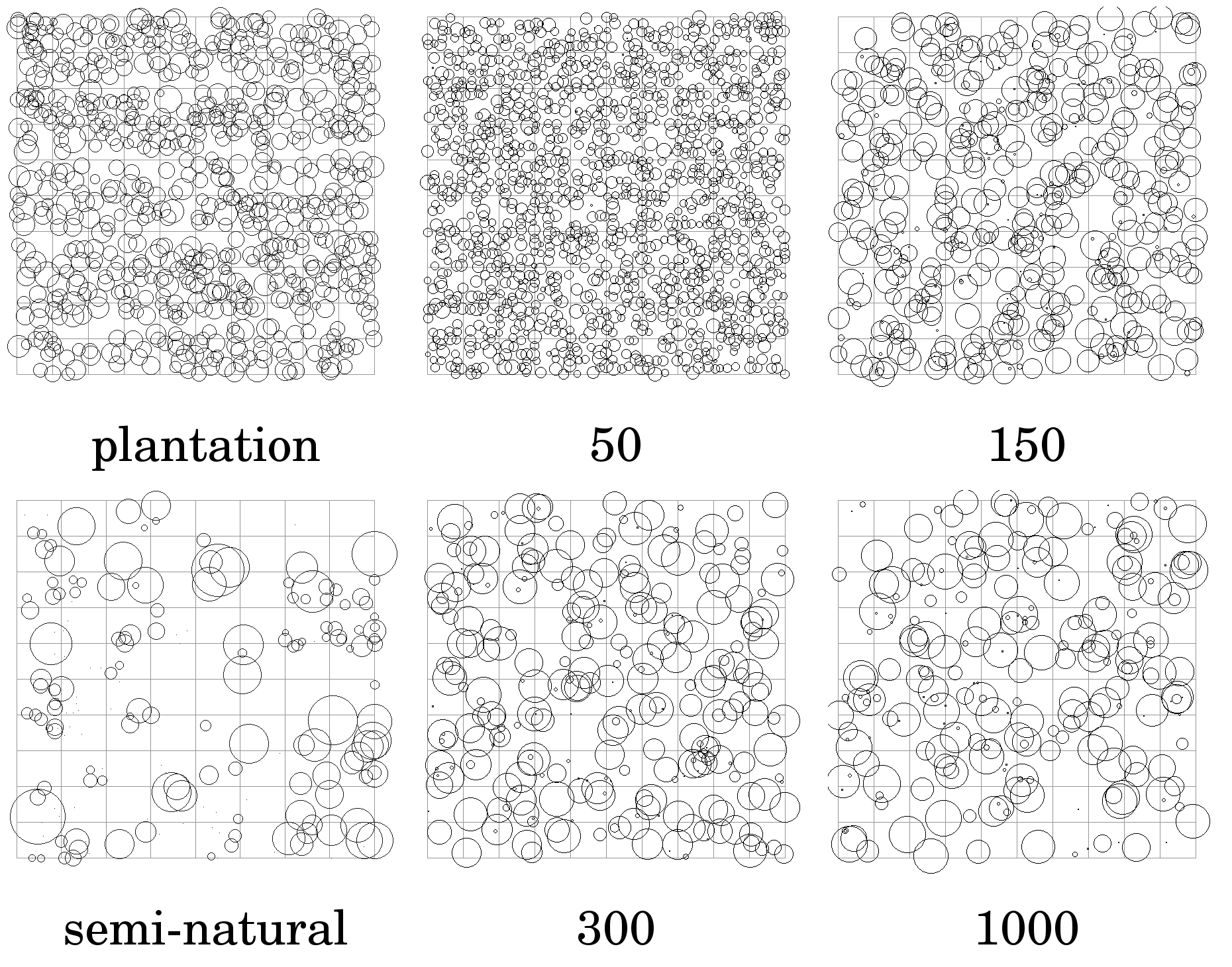}}
  \caption{Pictorial representation of 1ha Scots Pine forest. Field
    data (Highland, Scotland, data from Forest Research, left column):
    78 year old plantation in Glenmore, semi-natural stand in Glen
    Affric. Simulated data (centre and right columns) at 50, 150, 300
    and 1000 years from planting. The diameter of each circle is
    proportional to the size (dbh) of the tree.}
  \label{fig_initialdev}
\end{figure}

\begin{figure}
  \centering
  \scalebox{0.70}[0.70]{\includegraphics{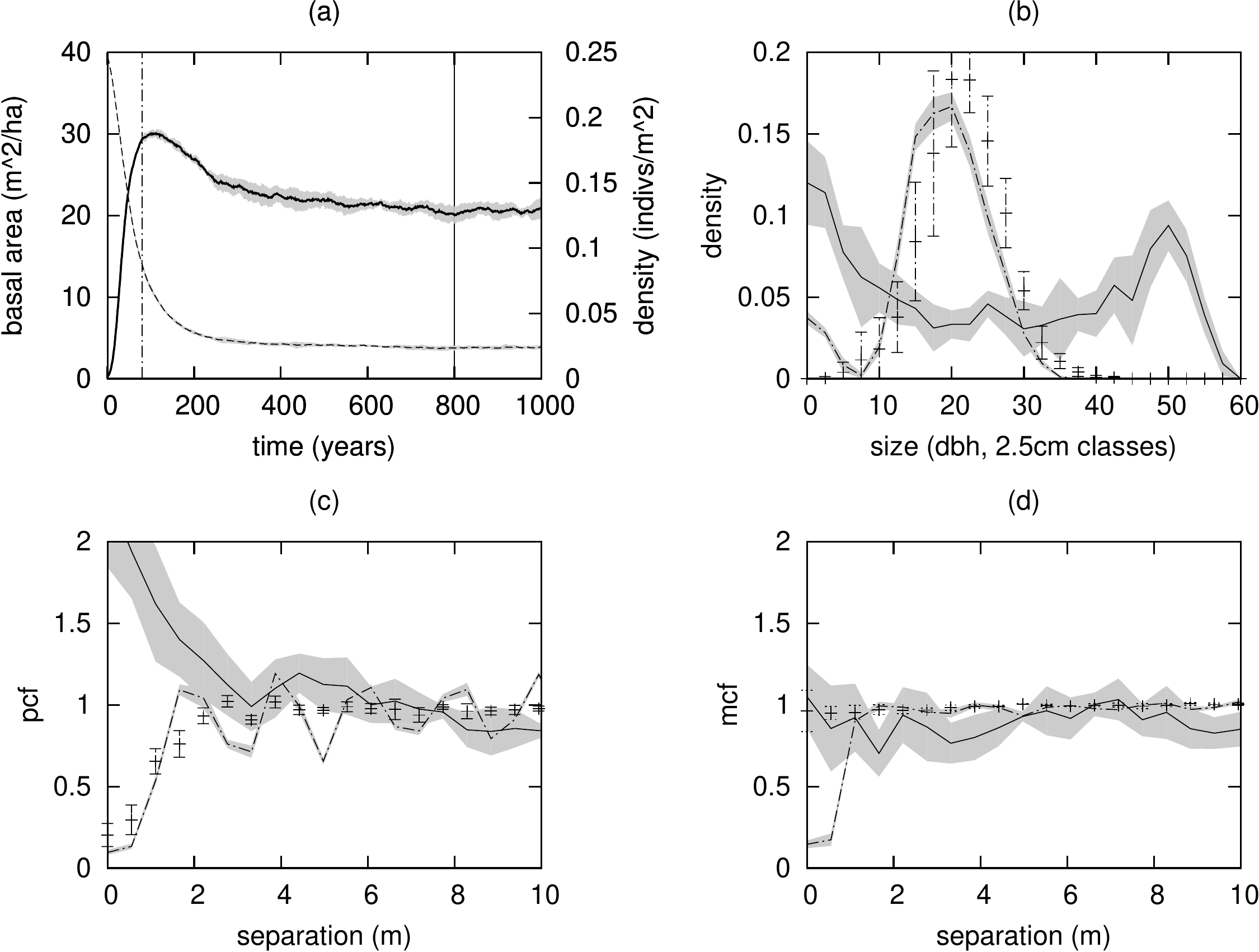}}
  \caption{The transition from plantation to steady state: development
    of key metrics through time, based on parameters in Table
    \ref{tab_parametertable}. Mean simulation results are represented
    by lines within a grey envelope (standard deviation).  (a)
    Evolution of density (dashed) and stand basal area (solid line),
    averaged over 10 simulations of a 1ha plot. (b) Size distribution
    at 80 (dash-dot) and 800 (solid) years. Points with error bars
    show the mean and standard deviation of 6 Glenmore plantation
    stands (78 years old). Steady-state comparison with natural
    stands is shown in Figure \ref{fig_seminat_match}.  (c) Pair
    correlation function -- time/colour as (b). (d) Mark correlation
    function -- time/line style as (b).}
  \label{fig_evolution}
\end{figure}

\subsubsection*{Plantation stage (``stand initiation'')}

The plantation structure initiated by forest management has a higher
density than a natural self-regenerating forest. We define this
transient stage of development as the period from time zero to the
point at which basal area initially peaks.  Reproduction is low, due
to individuals' small size.  Density is thus dominated by mortality,
and falls rapidly. Competition is also relatively low, meaning that
individuals can express the majority of their potential growth. Basal
area increases rapidly as a consequence (see Figure
2).

Our simulated density and size distribution of the model are fairly
close to those of plantation stands at Glenmore (simulated at 80 years
vs. dataset: 0.09063 vs 0.08523 individuals per m$^{2}$).  However,
basal area is notably underestimated (29.22 vs 36.67m$^{2}$ha$^{-1}$).
The reason for this is apparent in Figure 2b; the
growth parameters estimated from semi-natural data alone give too slow
growth in the simulated population (the modal size at 80 years is
lower). There are also more very small individuals in the simulation
at this stage.  This may indicate problems with the recruitment
process in the model, or be related to poor deer control at the
Glenmore plantations.

Stochastic variation in growth, and asymmetric competition, lead to an
increase in the spread of sizes of individuals (the initial size
distribution is a delta peak at $s=0.01$m). Size asymmetry is often
cited as a key driving force in plant community dynamics
\citep{adams07,perry03,weiner01}. In our model, competitive size
asymmetry is the primary factor affecting the variance (spread) of the
size distribution during the plantation stage: it is almost
independent of any other parameter, or even starting spatial
configuration (see Table 3, Appendix 2, ESM).

Low reproduction means that spatial structure is governed by the
starting configuration. The pair correlation function (PCF), giving
the relative density of pairs of individuals with given separation
\citep{penttinen92}, clearly shows the signature of the lattice during
this stage (Figure 2c, 80 years -- peaks are at
multiples of the lattice spacing). In field data, however, the
lattice pattern is less clearly defined: this is a data collection
issue, individuals' locations were measured to an accuracy of one
metre. However, the PCF displays the same short range inhibition as
the data.

The mark correlation function (MCF) measures the relative size of
individuals forming pairs at a given separation, compared to the
global average \citep{penttinen92}. Figure 2d
suggests that the average size of pairs at short ranges (less than 2m
separation) in the simulation is inhibited.  However, this feature is
an artefact of regeneration seen in the simulation (that is, the peak
of small individuals discussed above) that is not present in the data.
Recomputing the mark correlation function ignoring these individuals
recovers the structure seen in the data (not shown) -- interaction has
less differential effect on growth of older individuals.

\subsubsection*{Thinning stage (``stem exclusion/understory reinitiation'')}

The high basal area (and high competition) state generated during the
plantation stage means that individual growth becomes stunted, and
mortality rates are elevated. Basal area thus reaches a peak. Removal
(``stem exclusion'') of suppressed (competitively inhibited)
individuals occurs, opening gaps in the canopy. This allows more
substantial regeneration to occur (gaps heighten $\mathbb{P}_{e}$ for
many of the potential offspring, high basal area ensures a large seed
source -- ``understory reinitiation''). The initial regular structure
is erased during this period, through mortality, regeneration and
differential growth. This change is apparent in both spatial
correlation functions (not shown), and in maps of the stand at 300
years (Figure 1).

This transitional period (from peak basal area to meta-stable state)
is around 4-5 times the length of the plantation stage. This is
notable; if ``old-growth'' refers to the long-run meta-stable state,
the model suggests that this is slower to attain than is commonly
assumed. Indeed, \citet{oliver96} point out that due to external
catastrophic disturbances, true old growth is rarely reached, taking
up to 1000 years to attain.

\subsubsection*{Long-run metastable state (``old-growth'')}
\label{sec_longrunstate}

In the long run, the model reaches a steady state where fecundity,
mortality and growth are in balance. Figure 2b
(solid line) shows the typical size structure present in the long run.
Only a small proportion of juveniles attain canopy size, but the
asymptotic nature of growth means that individuals accumulate in the
higher size classes.

The size distribution is stable. For a dense forest, this is evident
intuitively -- when a gap in the canopy opens, the smaller younger
trees are waiting to grow into it. Here, reductions in canopy density
reduce local interactions, and temporarily allow trees that have
stopped growing to increase in size, quickly refilling gaps.

Spatial structure displays a more irregular pattern than earlier
stages. The effect of individual interactions upon growth are evident
in the reduced size of adjacent pairs (Figure 2d),
whilst local dispersal of seedlings leads to a heightened PCF at short
ranges (Figure 2c).  Some authors
\citep[e.g.][]{barbeito08} have noted that regeneration sometimes
occurs in explicitly clustered patterns. This may be due to external
environmental factors, but lead to a similar observed PCF.

\subsubsection*{Comparison with ``semi-natural'' data}
\label{sec_seminat}

We would like to identify whether the long run steady state behaviour
of the model mimics that of a real forest. The basic numbers appear
roughly correct (simulated vs data: density - 0.0327 vs 0.0165-0.025
individuals per m$^{2}$; basal area - 23.9 vs
18.6-25.2m$^{2}$ha$^{-1}$). However, all available data is from
``semi-natural'', as opposed to ``equilibrium'' stands. Whilst
relatively untouched over the last 90 years, these stands have been
managed in the past, their current state reflecting these historic
interventions.

The most recent management of these stands was the removal of the
strongest trees to assist with the war effort (ending in 1918).  For
the Rannoch stands, their state at this time may (in part) be deduced
using the individual growth/age data from the annual increment cores
discussed in Section \ref{sec_paramsec}. Working back from the current
(actual) diameter, the size of each tree at 1918 can be estimated, and
consequently the total basal area (in 1918) of the trees still present
today (plot 4: 9.2m$^{2}$, plot 6: 2.5m$^{2}$).  There is known to
have been low mortality in these stands over this period (the study
plots were established in 1948).

To simulate this, the model equilibrium state is thinned to
$10$m$^{2}$ basal area, by removing trees from the largest $60\%$. The
model is then run for a further 80 years before comparison (Figure
3).

The data stands display high variability, reflecting the effect of
site-specific processes and previous management on each site. Similar
signatures are seen, however: the PCF (Figure
3a) shows clustering of individuals in all
stands, whilst the MCF (Figure 3b) displays
inhibition of growth/size at short ranges. The PCF of the simulation
displays the same form as the data.  However, the MCF generated by the
model appears too homogeneous. Glen Affic displays the strongest size
inhibition at short range (lowest MCF), but also has the largest
relative density of small trees (and larger than that of the model).
Interestingly, the MCF for Glen Affric omitting juveniles ($<0.1$m
dbh) does not display any significant inhibition at short ranges (not
shown). The implication is that interaction affects diameter growth of
juveniles more than that of mature trees (not incorporated into the
model).

Size distribution of the data is generally characterised by a wide
spread and a ``canopy'' peak at a moderate size (Figure
3c). This is similar to the simulation output,
but there are two important issues. Firstly, the estimated growth
parameters (Appendix 1, ESM) limit the asymptotic
size at $\exp(\alpha/\beta)=62.9$cm, meaning that the few very large
trees observed in data cannot be created by the model. Secondly, the
variation in size in the data is not directly related to variation in
age, as it is in the model (Figure 3d). Both
issues are tackled below.

\begin{figure}
  \centering
  \scalebox{0.70}[0.70]{\includegraphics{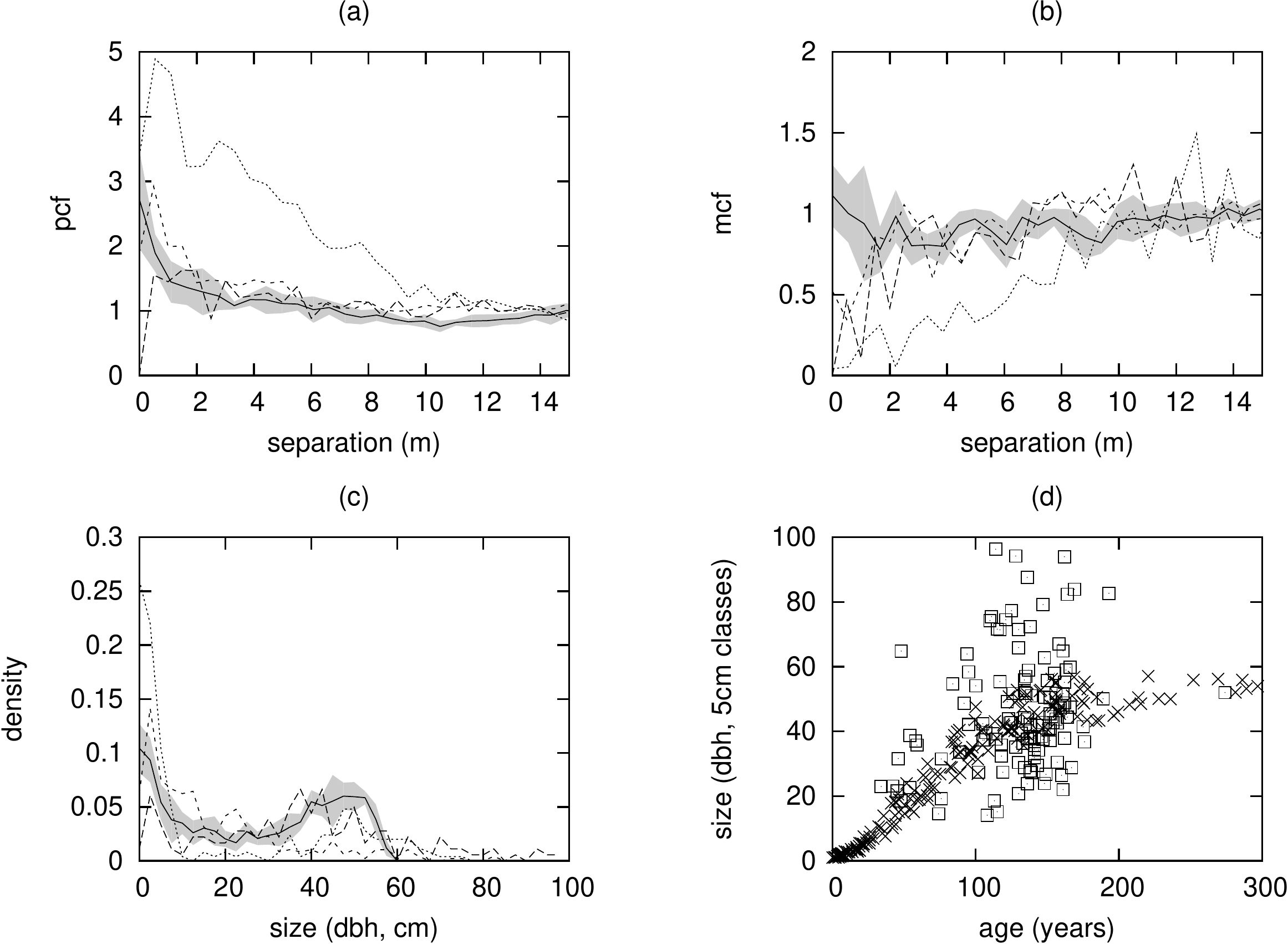}}
  \caption{Comparing statistics from ``semi-natural'' datasets with
    simulation output. The data stands were heavily managed prior to
    1918; we approximate this by running a model stand to equilibrium,
    thinning to 10m$^{2}$ basal area (removing trees randomly from the
    largest $60\%$), and running for a further 80 years (see Section
    \ref{sec_seminat}). Solid line and grey envelopes in (a),(b) and
    (c) are simulated results. Data: Rannoch plot 4 (dashed), Rannoch
    7 (fine dash), Glen Affric (dotted). Spatial correlation functions
    display a similar signature for all stands - clustering of
    individuals (a), and inhibition of growth/size at short ranges
    (though this is not seen to a great extent in simulation, see main
    text) (b). Size distribution (c) varies between the stands,
    reflecting the management history. (d) shows the variability in
    size attained at a given age present in the data (individual trees
    represented by $\boxdot$ -- data only available for Rannoch
    stands),compared with the simulation 80 years after the
    intervention ($\times$).}
  \label{fig_seminat_match}
\end{figure}

\subsubsection*{Size vs age - random asymptotic size}

There are two possible causes of the discrepancy between the model and
size and size/age distributions. Firstly, the difference may be induced
by the model, in its characterisation of competition and growth.
Secondly, it may be due to intrinsic or environmental variation
between the growth of the trees in real stands.

To address these issues we first explored increasing the strength of
competition, by increasing $\gamma$. This increases variability in
modelled growth histories, but at early times leads to unrealistic
size distribtions compared with plantation data (results not shown,
but see Appendix 2, ESM, for generalisations of behaviour).  An
alternative hypothesis was that observed growth variation could be
accounted for by competition ``accumulating'' throughout an
individual's life, causing a permanent adjustment to its asymptotic
size. Unfortunately, this does not provide a greatly improved
explanation of the data either (in NLME analysis, despite an
improvement in fit, parameter standard deviations are not reduced --
see Appendix 1, ESM). Furthermore, model behaviour is not altered
significantly without increasing $\gamma$ from the estimated value, as
above (which is again inconsistent with plantation data).

NLME analyses of simulated data (where simulated individuals have
identical parameter values) recovers the growth parameters used
accurately and with low standard deviation, across a range of
scenarios (not shown). This contrasts with analysis of observed data
(Appendix 1, ESM), suggesting the existence of genuine variation
between individuals in the data stands, or small scale environmental
variation.

Therefore, rather than use the mean values of growth parameters
obtained from the NLME analysis as Section \ref{sec_paramsec}, we also
perfomed simulations selecting $\alpha$,$\beta$ from the bivariate
Normal distribution estimated by that analysis (with estimated
correlation $\rho=0.988$). This obtains accurate steady state
behaviour, but there is excessive size variance at age 80 years. The
variability inferred from the semi-natural data is inconsistent with
the plantation data, where initial growth rate is relatively uniform
across individuals.

In Equation \ref{eqn_gomp}, $\alpha$ controls the initial growth rate,
while $\alpha/\beta$ determines asymptotic size. Thus, a second
approach was devised: fixing $\alpha$ at the mean from the NLME
analysis, whilst drawing $\exp(\alpha/\beta)$ (asymptotic size) from
the observed sizes of individuals greater than 100 years old in the
data stands.  This obtains a much better overall match with available
data than other methods (including the joint age-size distribution,
see Figure 4) and avoids unrealistic maximum tree
size as seen previously. It can also be sampled directly from a target
stand.

\begin{figure}
  \centering
  \scalebox{0.70}[0.70]{\includegraphics{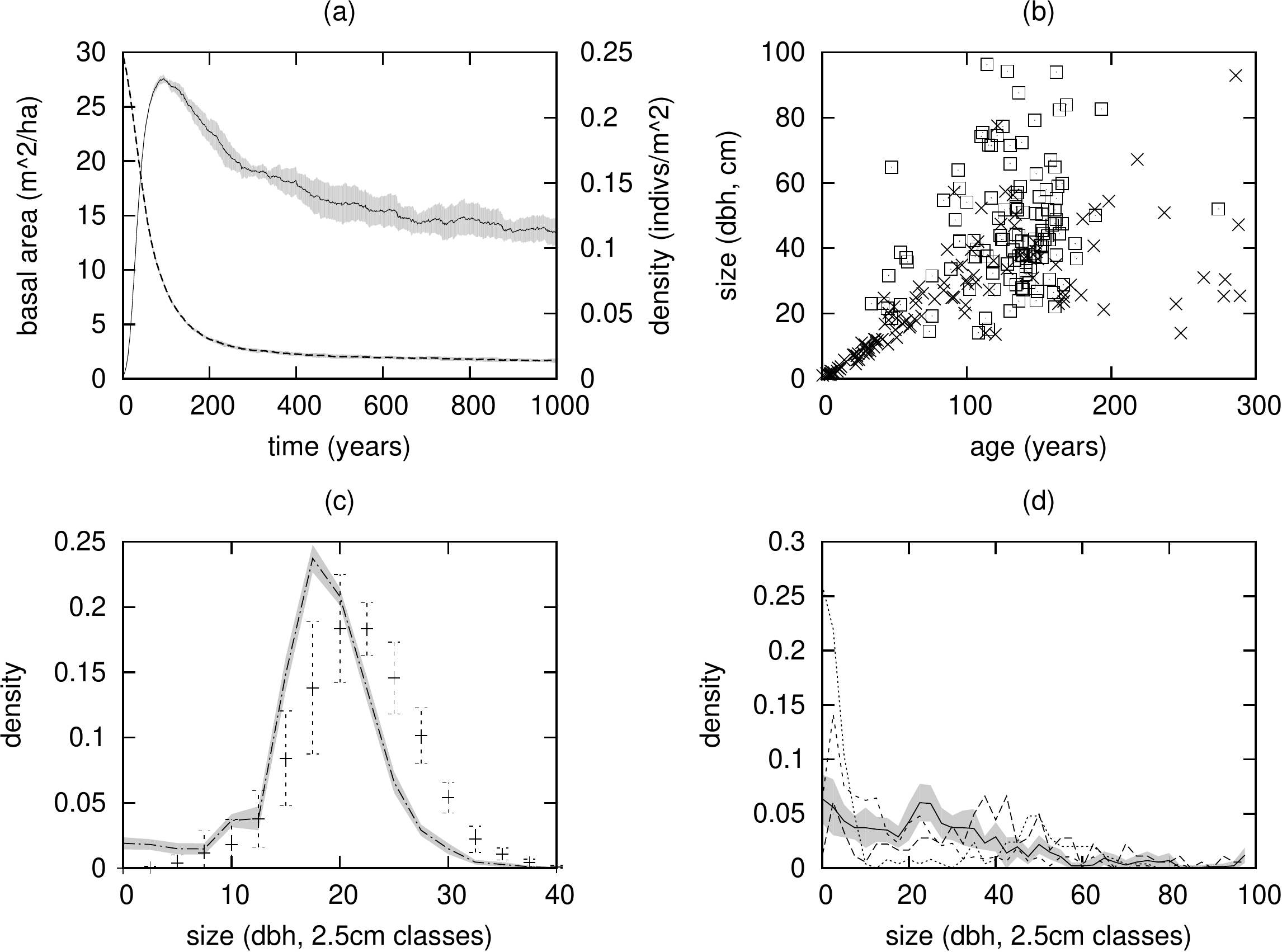}}
  \caption{Results obtained from populations with fixed $\alpha$,
    sampling $\exp(\alpha/\beta)$ from observed sizes of individuals
    greater than 100 years old at 1990 in Rannoch plots.
    \emph{Corresponding Figures for the non-random model are given in
      italics.} Again, simulation means are represented by lines
    within a standard deviation envelope. (a) density (dashed) and
    basal area (solid) (\emph{Figure \ref{fig_evolution}a}).
    Comparison with real stand data: (b) size versus age 80 years
    after the intervention described in Section \ref{sec_seminat},
    compared with the Rannoch plot (\emph{Figure
      \ref{fig_seminat_match}d}.  (c) size distribution at 80 years
    (line) versus Glenmore planation average and standard deviation
    (error bars) (\emph{Figure \ref{fig_evolution}b}). (d) size
    distribution at 880 years (solid line) versus Rannoch 4 (dashed
    line), Rannoch 7 (fine dash), and Glen Affric data (dotted)
    (\emph{Figure \ref{fig_seminat_match}c}).}
  \label{fig_randommodel}
\end{figure}

\subsubsection{Summary}

We have constructed a simple model that, when parameterised from
observed data, matches fairly well the qualitative and quantitative
behaviour of real forests on a stand level, despite large
uncertainties in management and environmental history for the
available data stands.  This model will now be used to demonstrate the
effectiveness of simple thinning strategies in accelerating the
transition to old growth.  Simulations in Section \ref{sec_manage} use
the homogeneous growth model for clarity; analogous results using the
model incorporating individual variation are presented in Appendix 3,
ESM.

\subsection{Acceleration of transition to old-growth state}
\label{sec_manage}

Stands possessing appealing characteristics are not necessarily
naturally formed \citep[see e.g.][]{edwards04}. Is the ``sustainable
irregular condition'' \citep{schutz01} the same as the old-growth
state described here? The expected qualities are
\citep{malcolm01,mason07}:
\begin{itemize}
\item {Full representation across the size classes with high variance
    in canopy size}
\item {Non-regular spatial distribution}
\item {High recruitment}
\end{itemize}
These conditions are met by the long term state of the model. What are
the main factors in achieving such a state? Natural regeneration is
key, and can be encouraged by thinning the existing canopy. Basal area
can be reduced \citep{hale01,edwards06a}, though creation of open
space (i.e. gaps) is also likely to be useful for light demanding
species such as Scots Pine.
 
We investigate thinning treatments applied to a mature plantation
intended to bypass or escape the period of unnaturally high basal
area, remove the lattice spatial pattern, and create suitable
conditions for the generation of a high-variance size structure.

Thinnings are often made on individuals' size relative to neighbours
and other members of the stand \citep[e.g.][]{edwards04}. For
simplicity these ``size-selection'' thinnings alone are presented
here; future work will compare these with spatially correlated (e.g.
patch) and interaction-selection (removing neighbours of selected
trees) thinnings. The presented results are however robust to
variation in size range and the use of spatial criteria instead of
size.

A single thinning has minimal effect on the subsequent stand dynamics.
However, multiple thinnings can alter the dynamics significantly. A
target basal area for the stand determines the extent of thinning,
although this may not be attained with a single treatment. Thinnings
to a target basal area of $18$m$^{2}$, repeated 5 times with an
interval of 2, 5 or 10 years are presented in Figure 5.

The basal area reduction due to thinning is temporary.  Thinning
releases some of the remaining large individuals from interaction
stress, which suddenly grow quickly. Widely separated interventions
have a more significant effect on the evolution of basal area than
those made in rapid succession (Figure 5a). However, the
initial boom in basal area above that of the steady state seems
unavoidable.

Basal area does not tell a complete story, however. Stand density
after the treatments shown approaches and stays close to the steady
state density (Figure 5b). 

Figure 5c compares the average size distribution at 200 for
each stand in 5a with the long run average steady state
size distribution. Management causes a clear reduction in canopy
density, and increases the average size of the canopy trees.  As the
time between treatments increases, the overall canopy density falls,
and the number of trees at the very largest sizes slightly increases,
bringing the stand closer to an old growth state.

Spatial structure is also improved; Figure 5d compares the
PCF at 200 years obtained by the thinning regimes described with that
of an unmanaged forest at 200 years (dotted), and its steady state
(thick solid). Despite the non-spatial thinning, the PCF is closer to
the steady state after this short time.  Spatially structured thinning
may further assist in generating specific patterns.

These results hold when fixing management interval and varying instead
the number of interventions, and also in thinning regimes approaching
the target basal area gradually -- it is the overall length of
management that is important.

\begin{figure}
  \centering
  \scalebox{0.70}[0.70]{\includegraphics{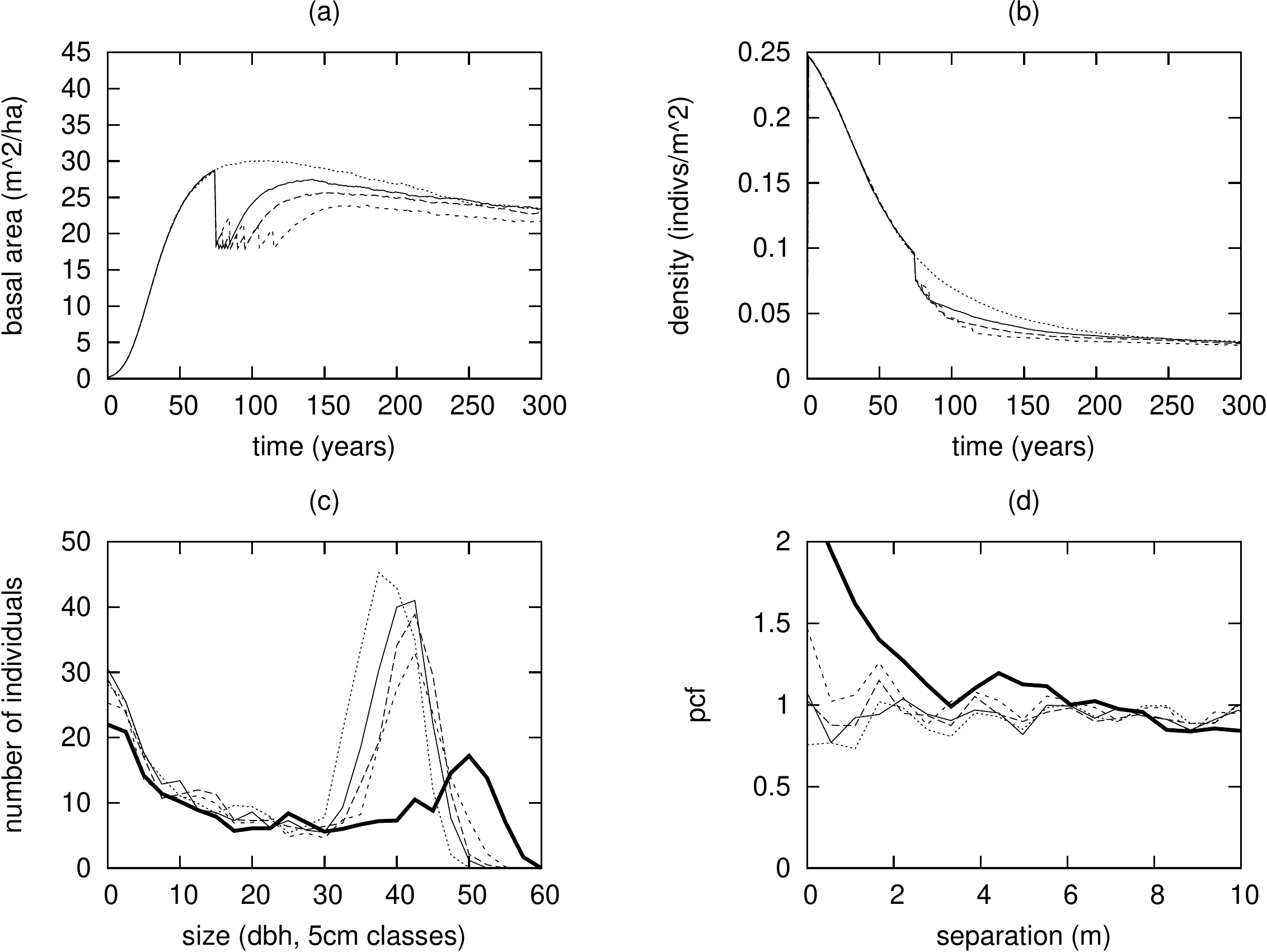}}
  \caption{Thinning randomly from 60-100\% of the size distribution,
    but altering the interval between treatments (again, 5 treatments
    starting at 80 years, with a target basal area of
    $=18$m$^{2}$ha$^{-1}$).Intervals: 2 years (solid), 5 years
    (dash), 10 years (fine dash).  Again, the effect on
    dynamics is demonstated by (a) basal area (b) density (c) size
    distribution at 200 years (d) pcf at 200 years. Dotted
    lines show the dynamics of an unmanaged forest, whilst the thick
    solid lines in (c) and (d) show the long-run steady state.}
  \label{thin2}
\end{figure}

\section{Discussion}

We have presented a model for the assembly of a single species forest
community incorporating both size and spatial structure.  Although
simple, our model can be parameterised to give a good qualitative
agreement with data for both plantation and semi-natural Scots Pine
forests that are geographically widely separated.

The model depends on parameters describing maximum size, growth,
competition and death.  While the growth and death parameters could be
taken as constant for all trees, it proved necessary for the maximum
size parameter to be drawn from a distibution.  This may represent
either genetic diversity or a variation in the ability of a given
location to support a tree.

The structure of simulated forests is strongly dependent on the
initial conditions, even after hundreds of years. The long-time
equilibrium state of the model has rather low density, with a highly
varied size (diameter) distribution. It appears to be stable, with no
evidence of cyclical variation in structural characteristics.

We applied the model to determine whether thinning treatment can be
effective in bringing a forest from a plantation to this steady state
more effectively than natural (unmanaged) regeneration.

Our work suggests that management \emph{can} have a clear effect on
the dynamics of a forest starting from a plantation state. Figure 5
shows that the effects of management are not insignificant;
first-order properties, and size/spatial structure can be manipulated
somewhat favourably. More complex planting and thinning regimes than
those considered here can also be implemented, though initial tests
suggest high levels of specificity are required to have the desired
effect on community structure.  Overall, there would appear to be no
direct ``short-cut'' to the old-growth state.

There are fundamental reasons for this. Firstly, the stand must be old
enough to have very large and mature trees, which are a central
component of the desired habitat. This rules out attainment of the
steady state size distribution after only one or two centuries.

Secondly, a young plantation is in a very low recruitment state. This
means that the community consists of individuals from a single age
cohort, which are either in the canopy or are very suppressed. Even
mediated by a planting scheme, this initial size and age structure
will persist in the community for some time.

Spatial structure on the other hand can be manipulated over short
timescales. A particular pattern of trees may be obtained quickly and
easily by selective thinnings. This also alters development of size
structure, by releasing selected individuals from competition, and
providing large gaps in which regeneration may readily occur.
Creating an old growth type state is also difficult with selective
planting: the trouble here is that one cannot predict which trees will
flourish into maturity.

While our emphasis has been on obtaining the steady state, the model
can also be used to investigate the management strategies needed to
achieve other aims such as maximum production of wood, of mature
trees, or even removal of CO$_2$ from the atmosphere.

To summarise, we find a relatively simple model can capture the main
features observed in the dynamics of a single-species Scots Pine
woodland at various developmental stages.  We applied the model to
investigate the attainment of an equilibrium ``old-growth'' state.
Whilst there appears to be no management strategy which will bring a
plantation rapidly to the natural state, structure can be manipulated
favourably over shorter timescales.

\section*{Acknowledgments}

We would like to acknowledge support from the Scottish Government and
the EPSRC funded NANIA network (grants GRT11777 and GRT11753). This
work made use of resources provided by the Edinburgh Compute
and Data Facility (ECDF) ( http://www.ecdf.ed.ac.uk/). The ECDF is
partially supported by the eDIKT initiative (http://www.edikt.org.uk).

\newpage
\appendix

\section{Growth parameter estimates}
\label{app_nlme_est}
Growth parameters were estimated from increment core data (radial
sections providing measurements of annual diameter growth over the
lifespan of each tree, taken at 1.0m height) from several semi-natural
Scots Pine stands in the Black Wood of Rannoch.  Parameters were
estimated from individual data taken from plots ``4'' and ``6'' (5 and 7
have less well known management history). To ensure estimation based
upon known competitive neighbourhoods, those individuals less than
than 10m from the plot boundary were excluded.  Furthermore, only
increments applying to growth after 1918 were used, this being the
date after which management (and consequently the state of the
community) is known with sufficient accuracy.

\emph{NLS} is a non-linear least squares fitting tool in R
\citep{rproject}, here applied to the complete set of increment
measurements. The fit computed is equivalent to assuming a single
growth curve generated all data points, which are regarded as
independent. \emph{NLME} is another tool in R, computing a non-linear
mixed efects model \citep{nlme}. This approach goes a step further, in
computing a nls fit for each individual in the poulation separately
(that is, hypothesised individual growth curves). This explicitly
estimates the variability present in the population by computing the
mean (the ``fixed effect'') and standard deviation (the ``random
effect'') of each parameter, and the correlation between them.

The precise definitions of the three models being fitted are:
\begin{align}
    growth =& size \times \left( \alpha - \beta \mbox{log}(size) \right)
    &\mbox{``no competition''}\\
    growth =& size \times \left( \alpha - \beta \mbox{log}(size) -
    \gamma \Phi \right) &\mbox{``competition''}\\
    growth =& size \times \left( \alpha - \beta \mbox{log}(size) -
    \gamma \sum_{t_{0} < t' < t} \Phi (t') \right) &\mbox{``cumulative competition''}
\end{align}

Residual Standard Error (RSE) summarises the difference between
observed and estimated values in the model ($RSE =
\sqrt{\mathbb{V}/n}$ where $\mathbb{V}$ is the variance of the
residuals and $n$ is the number of observations). Aikake's An
Information Criterion (AIC, \citet{aikake74}) is a likelihood-based
measure with a penalisation related to the number of model parameters
$k$: $AIC = -2 log(L)-2k$. A lower value indicates a more parsimonious
model.

Given the structure of the data (subsets of the complete data describe
the growth curves of individual trees), the NLME approach is
conceptually more appropriate, a point confirmed by the uniformly
lower RSE and AIC for the NLME models. That different numbers of
measurements are available for different trees (depending on their
age) makes this all the more important. It transpires that there is
rather large variation in growth rates, that cannot be described by a
fixed set of parameters across the population. In the NLME analysis,
the computed standard deviation for each parameter is on the same
order as the mean, and in the case of $\gamma$, is actually larger.
$\alpha$ and $\beta$ were found to be strongly correlated (in the
``competition'' model, $\rho_{\alpha,\beta}=0.988$,
$\rho_{\alpha,\gamma}=0.557$, $\rho_{\beta,\gamma}=0.481$).

Despite the improved fit offered by the cumulative competition model,
the basic competition model was selected for analysis and simulation
due to its lack of dependence upon history (maintaining the Markov
property of the process). It is also important to realise that
spatio-temporal data of the type provided by these increment cores are
much more laborious to collect, and as a consequence far less widely
available, than the marked point process (single point in time) data
that are usually used in spatial analyses.

\begin{table}[h]
  \centering
  \caption{Estimated parameters for non-linear growth models fitted to data from Rannoch plots 4 and 6 combined (plot 5 and 7 omitted due to missing recent management history; growth curves computed based upon increments after 1918 for individuals further than 10m from an edge). Function fitted: Gompertz with and
    without competition term (interaction formulated as in model
    description with parameters shown).}
  \vspace{0.1in}
  \begin{tabular}{c|ccc|cccc}
    & \multicolumn{3}{c|}{nls} &\multicolumn{4}{c}{nlme} \\
    \hline 
    & LS Estimate & RSE & AIC & Fixed ($\mu$) & Random ($\sigma$) & RSE & AIC \\
    \hline
    \multicolumn{8}{l}{{\bf no competition}}\\
    \hline
    $\alpha$ & 0.0426 &  0.311 &3256.9 & 0.132 & 0.0931 &0.117 & -8141.2\\
    $\beta$ & 0.00909 & & & 0.0359 & 0.0281 && \\    
    \hline
    \multicolumn{8}{l}{{\bf competition}}\\
    \hline
    $\alpha$ &0.0828 &0.269 & 1369.8 &1.308&
    0.103 & 0.116 & -8194.0 \\
    $\beta$ &0.0177 & & & 0.0318 &0.0286 &&\\
    $\gamma$ & 4.46e-05 & & & 6.51e-05 & 6.97e-05 &&\\
    \hline
    \multicolumn{8}{l}{{\bf cumulative competition}}\\
    \hline
    $\alpha$ & 0.0684 & 0.275 & 1646.8 & 0.146 &
    0.0967 & 0.115 & -8251.4 \\
    $\beta$ & 0.0146 & & & 0.0410 & 0.0310 &&\\
    $\gamma$ & 4.56e-07 & & & -7.17e-07& 1.07e-06 &&\\

  \end{tabular}
  \vspace{0.1in}
  \label{tab_growth_param_est}
\end{table}
\newpage

\section{Effects of parameter variation}
\label{app_paramvar}

This appendix provides a brief summary of the effects of parameter
variation upon various aspects of model behaviour. Model behaviour is
robust: the effects desribed hold for at least an order of magnitude
above and below the parameters used in the main text (Table 1 in main
text), unless otherwise stated. The thinning stage is not included
here; it is a transient state with properties dependent upon the
relative properties of the plantation and steady-state under the
chosen parameterisation.

\subsection{Plantation stage}
Largely speaking, changes to individual parameters have predictable
effects upon the properties of the community's early development.
However, there are some counter-intuitive effects. For example,
increasing the effect of interaction upon mortality ($\mu_{2}$)
increases the mean size at 80 years, through density reduction and a
corresponding decrease in suppression of growth rate.

Increasing ``mortality'' in Table \ref{tab_parameffect1} refers to
increasing both $\mu_{1}$ and $\mu_{2}$ whilst fixing their ratio,
ensuring that baseline and interaction induced mortality always have
the same relative strength.

The variance of the size distribution at 80 years appears almost
unaffected by any parameter, except the degree of size asymmetry in
the interaction kernel ($k_{s}$). Interestingly, the only statistics
considered here (including long-run behaviour) that are affected by
$k_{s}$ relate to the shape of the plantation/early stage size
distribution.

\begin{table}[h]
  \centering
  \caption{A summary of the qualitative effect on plantation development (as summarised by
    various statistics) of increasing any parameter of the model in
    isolation. In columns, $\rho$ is density and $s$ is size, with
    subscripts refering to time, $\mathbb{E}$ and  $\mathbb{V}$ to
    expected value and variance. $BA_{peak}$ is the maximum basal area
    attained by the population, $t_{BA peak}$ the time at which it occurs
    Increasing ``mortality'' refers to increasing $\mu_{1}$ and
    $\mu_{2}$ whilst fixing their ratio, and increasing ``growth''
    means increasing both $\alpha$ and $\beta$, whilst fixing their
    ratio.}
  \vspace{0.1in}
  \begin{tabular}{c|cc|cc|cc}
    & \multicolumn{6}{l}{Statistic (plantation)}\\
    Parameter & $\rho_{80}$ & $BA_{80}$ & $\mathbb{E}(s_{80})$ & $\mathbb{V}(s_{80})$ & $BA_{peak}$ & $t_{BA peak}$ \\
    \hline
    {\bf rates}&&&&&&\\
    $f$ & $+$ & $+$ & 0 & 0 & $-$ & $-$ \\
    mortality & $-$ & $-$ & + & 0 & 0 & 0 \\
    growth & $+$ & $+$ & $+$ & 0 & $+$ & $-$ \\
    {\bf interaction}&&&&&&\\
    $\mu_{2}$ & $-$ & $-$ & $+$ & 0 & $-$ & $-$ \\
    $\gamma$ & $-$ & $-$ & $-$ & 0 & $-$ & $-$ \\
    {\bf kernels}&&&&&&\\
    $k_{d}$ & $+$ & $+$ & $+$ & 0 & $+$ & 0 \\
    $k_{s}$ & 0 & 0 & $-$ & $+$ & 0 & 0 \\
  \end{tabular}
  \vspace{0.1in}
  \label{tab_parameffect1}
\end{table}
\newpage

\subsection{Old-growth stage}
Turning to longer-run behaviour, within the parameter space presented,
steady state density and basal area are increased by increasing
fecundity or growth speed, or decreasing mortality (all other
parameters remaining equal, results shown in Table
\ref{tab_parameffect2}). Interestingly, decreasing mortality further
to unrealistically low levels leads to a decrease in steady state
basal area. This somewhat surprising result occurs due to individual
growth being highly limited by density, rather than by lifespan
(results not shown).

Fixing population dynamic rates whilst altering the interaction
multipliers and kernels also has an effect on behaviour. Increasing
$\mu_{2}$ leads to a lower density, but greater sized, canopy. It also
reduces the size of close pairs (lower MCF). Increasing the effect of
interaction on growth ($\gamma$) reduces the density and size of the
canopy, whilst also causing a reduction in the size (but not density)
of close pairs (lower MCF).

The effects of $k_{d}$ are similar to the plantation case. $k_{s}$ has
no noticable effect on any aspect of long-run behaviour.

\begin{table}[h]
  \centering
  \caption{The effect on steady state behaviour (as summarised by
    various statistics) of increasing any parameter of the model in
    isolation. Again, $\rho$ is density. $s_{canopy}$ is the mean size
    of canopy trees. $\rho_{closepairs}$ is the
    value of the PCF at short ranges, whilst $s_{close
      pairs}$ is the value of the MCF at short ranges. Increasing
    ``mortality'' refers to increasing $\mu_{1}$ and
    $\mu_{2}$ whilst fixing their ratio, and increasing ``growth''
    means increasing both $\alpha$ and $\beta$, whilst fixing their
    ratio. Canopy density is relative to total density (proportion of
    individuals $>50\%$ of maximum size).}
  \vspace{0.1in}
  \begin{tabular}{c|cc|cc|cc}
    & \multicolumn{6}{l}{Statistic (steady state)}\\
    Parameter & $\rho$ & BA & $\rho_{canopy}$ & $s_{canopy}$ & $\rho_{close pairs}$ & $s_{close pairs}$\\
    \hline
    {\bf rates}&&&&&&\\
    $f$ & $+$ & $+$ & $+$ & $-$ & $+$ & $+$ \\
    mortality & $-$ & $-$ & $-$ & $+$ & $-$ & $-$ \\
    growth & $+$ & $+$ & $+$ & 0 & 0 & $+$ \\
    {\bf interaction}&&&&&&\\
    $\mu_{2}$ & $-$ & $-$ & $-$ & $+$ & 0 & $-$ \\
    $\gamma$ & 0 & $-$ & $-$ & $-$ & 0 & $-$ \\
    {\bf kernels}&&&&&&\\
    $k_{d}$ & $+$ & $+$ & $+$ & 0 & $-$ & 0 \\
    $k_{s}$ & 0 & 0 & 0 & 0 & 0 & 0 \\
  \end{tabular}
  \vspace{0.1in}
  \label{tab_parameffect2}
\end{table}
\newpage

\section{Management under variable growth}

This Section simply presents the same results relating to management
as those in Section 4 of the main text (``Acceleration of transition
to old-growth state''). The statistics computed using the model in
which individual variation (``Model 2'') is allowed show a similar but
slightly less clear pattern.

Temporal evolution of basal area and density show precisely the same
pattern as those under the homogeneous growth model (``Model 1'') --
the longer the duration of management, the closer they remain to the
steady state after thinning.

\begin{figure}
  \centering
  \scalebox{0.70}[0.70]{\includegraphics{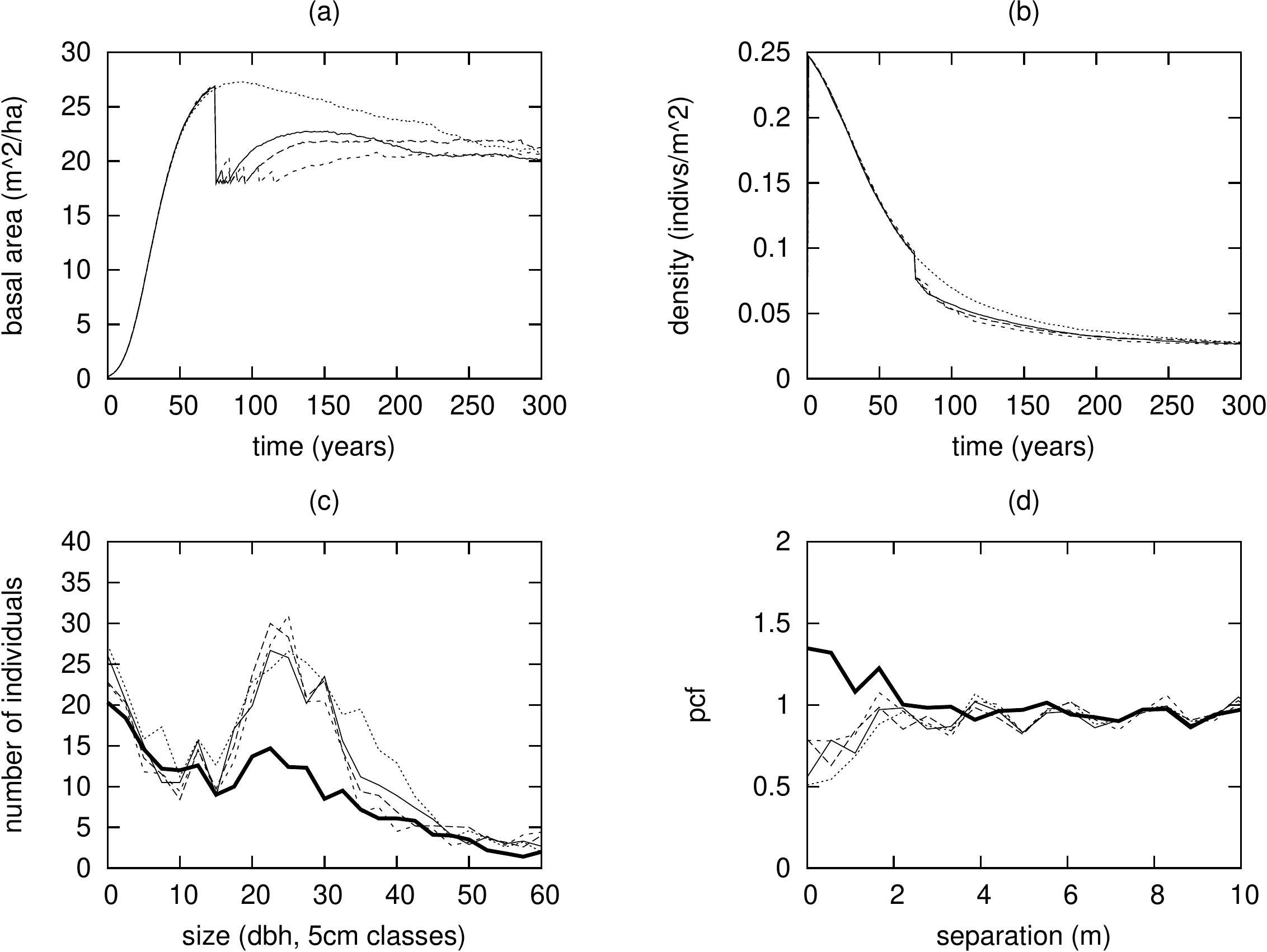}}
  \caption{Thinning randomly from 60-100\% of the size distribution,
    but altering the interval between treatments (again, 5 treatments
    stating at 75 years, with a target basal area of
    $=18$m$^{2}$ha$^{-1}$).Intervals: 2 years (solid), 5 years
    (dash), 10 years (fine dash).  Again, the effect on
    dynamics is demonstated by (a) basal area (b) density (c) size
    distribution at 200 years (d) pcf at 200 years. The dotted
    lines show the dynamics of an unmanaged forest, whilst the thick
    solid lines in (c) and (d) show the long-run steady state.}
  \label{fig_manageapp}
\end{figure}

Under Model 1, the size distribution demonstrated a shift in canopy
peak as the total duration of management increased, with a larger size
and lower density (Figure 5c in main text). Under Model 2, the size
distribution shows no increase in the size of trees in the canopy,
only a reduction in density towards that of the steady-state
distribution (Figure \ref{fig_manageapp}c here). This is due to the
much lower mean asymptotic size under Model 2.

With regards the pair correlation function (PCF), the shift towards
the steady state appears to be present but is also slightly less clear
-- the shift towards a clutered pattern being slower to occur under
Model 2 (Figure \ref{fig_manageapp}d here).

\begin{thebibliography}{}


\bibitem[Aikake, 1974]{aikake74}
Aikake, H. (1974).
\newblock A new look at the statistical model identification.
\newblock {\em IEEE Transactions on Automatic Control}, 19(6):716--723.

\bibitem[Pinheiro et~al., 2009]{nlme}
Pinheiro, J., Bates, D., DebRoy, S., Sarkar, D., and the R~Core~team (2009).
\newblock {\em nlme: Linear and Nonlinear Mixed Effects Models}.
\newblock R package version 3.1-92.

\bibitem[{R Development Core Team}, 2005]{rproject}
{R Development Core Team} (2005).
\newblock {\em R: A language and environment for statistical computing}.
\newblock R Foundation for Statistical Computing, Vienna, Austria.
\newblock {ISBN} 3-900051-07-0.


\bibitem[Adams et~al., 2007]{adams07}
Adams, T., Purves, D.~W., and Pacala, S.~W. (2007).
\newblock Understanding height-structured competition: is there an R* for
  light?
\newblock {\em Proceedings of The Royal Society B: Biological Sciences},
  274:3039--3047.

\bibitem[Barbeito et~al., 2008]{barbeito08}
Barbeito, I., Pardos, M., Calama, R., and Canellas, I. (2008).
\newblock Effect of stand structure on stone pine (pinus pinea l.) regeneration
  dynamics.
\newblock {\em Forestry}, 81(5):617--629.

\bibitem[Bolker and Pacala, 1997]{bolker97}
Bolker, B. and Pacala, S.~W. (1997).
\newblock Using moment equations to understand stochastically driven spatial
  pattern formation in ecological systems.
\newblock {\em Theoretical Population Biology}, 52:179--197.

\bibitem[Botkin et~al., 1972]{botkin72}
Botkin, D.~B., Janak, J.~T., and Wallis, J.~R. (1972).
\newblock Some ecological consequences of a computer model of forest growth.
\newblock {\em Journal of Ecology}, 60(3):849--872.

\bibitem[Busing and Mailly, 2002]{busing02}
Busing, R.~T. and Mailly, D. (2002).
\newblock Advances in spatial, individual-based modelling of forest dynamics.
\newblock {\em Journal of Vegetation Science}, 15:831--842.

\bibitem[Canham et~al., 2004]{canham04}
Canham, C.~D., LePage, P.~T., and Coates, K.~D. (2004).
\newblock A neighbourhood analysis of canopy tree competition: effects of
  shading versus crowding.
\newblock {\em Canadian Journal of Forest Research}, 34(4):778--787.

\bibitem[Clark et~al., 2004]{clark04}
Clark, J.~S., LaDeau, S., and Ibanez, I. (2004).
\newblock Fecundity of trees and the colonization-competition hypothesis.
\newblock {\em Ecological Monographs}, 74(3):415--442.

\bibitem[Comas, 2005]{comas05}
Comas, C. (2005).
\newblock {\em Modelling forest dynamics through the development of spatial and
  temporal marked point processes}.
\newblock PhD thesis, University of Strathclyde.

\bibitem[Cox and Miller, 1965]{cox65}
Cox, D. and Miller, H. (1965).
\newblock {\em The theory of stochastic processes}.
\newblock Methuen, London.

\bibitem[Edwards and Mason, 2004]{edwards04}
Edwards, C. and Mason, B. (2004).
\newblock Scots pine variable intensity thinning plots - glenmore.
\newblock Technical report, Forest Research.

\bibitem[Edwards and Mason, 2006]{edwards06b}
Edwards, C. and Mason, W. (2006).
\newblock Stand structure and dynamics of four native scots pine (pinus
  sylvestris l.) woodlands in northern scotland.
\newblock {\em Forestry}, 79(3):261--268.

\bibitem[Edwards and Rhodes, 2006]{edwards06a}
Edwards, C. and Rhodes, A. (2006).
\newblock The influence of ground disturbance on natural vegetation in a native
  pinewood: results after 60 years.
\newblock {\em Scottish Forestry}, 60:4--11.

\bibitem[Featherstone, 1998]{tfl98}
Featherstone, A.~W. (1998).
\newblock Species profile: Scots pine.
\newblock http://www.treesforlife.org.uk/tfl.scpine.html.
\newblock Date accessed: 27 July 2009.

\bibitem[{Forestry Commission}, 2009]{forestry09}
{Forestry Commission} (2009).
\newblock Scots pine - pinus sylvestris.
\newblock http://www.forestry.gov.uk/forestry/infd-5nlfap.
\newblock Date Accessed: 27 July 2009.

\bibitem[Gillespie, 1977]{gillespie77}
Gillespie, D.~T. (1977).
\newblock Exact stochastic simulation of coupled chemical reactions.
\newblock {\em The Journal of Physical Chemistry}, 81(25):2340--2361.

\bibitem[Gratzer et~al., 2004]{gratzer04}
Gratzer, G., Canham, C., Dieckmann, U., Fischer, A., Iwasa, Y., Law, R., Lexer,
  M.~J., Sandmann, H., Spies, T.~A., Splectna, B.~E., and Swagrzyk, J. (2004).
\newblock Spatio-temporal development of forests -- current trends in field
  methods and models.
\newblock {\em Oikos}, 107:3--15.

\bibitem[Hale, 2001]{hale01}
Hale, S.~E. (2001).
\newblock Light regime beneath sitka spruce plantations in northern britain:
  preliminary results.
\newblock {\em Forest Ecology and Management}, 151:61--66.

\bibitem[Larocque, 2002]{larocque02}
Larocque, G.~R. (2002).
\newblock Examining different concepts for the development of a
  distance-dependent competition model for red pine diameter growth using
  long-term stand data differing in initial stand density.
\newblock {\em Forest Science}, 48(1):24--34.

\bibitem[Law et~al., 2009]{law09}
Law, R., Illian, J., Burslem, D. F. R.~P., Gratzer, G., Gunatilleke, C. V.~S.,
  and Gunatilleke, I. A. U.~N. (2009).
\newblock Ecological information from spatial patterns of plants: insights from
  point process theory.
\newblock {\em Journal of Ecology}, 97:616--628.

\bibitem[Law et~al., 2003]{law03}
Law, R., Murrell, D.~J., and Dieckmann, U. (2003).
\newblock Population growth in space and time: spatial logistic equations.
\newblock {\em Ecology}, 84(1):252--262.

\bibitem[Lindstrom and Bates, 1990]{lindstrom90}
Lindstrom, M.~J. and Bates, D.~M. (1990).
\newblock Nonlinear mixed effects models for repeated measures data.
\newblock {\em Biometrics}, 46:673--687.

\bibitem[Loewenstein, 2005]{loewenstein05}
Loewenstein, E.~F. (2005).
\newblock Conversion of uniform broadleaf stands to an uneven-aged structure.
\newblock {\em Forest Ecology and Management}, 215:103--112.

\bibitem[Malcolm et~al., 2001]{malcolm01}
Malcolm, D., Mason, W., and Clarke, G. (2001).
\newblock The transformation of conifer forests in britain: regeneration, gap
  size and silvicultural systems.
\newblock {\em Forest Ecology and Management}, 151:7--23.

\bibitem[Mason et~al., 2007]{mason07}
Mason, W., Connolly, T., Pommerening, A., and Edwards, C. (2007).
\newblock Spatial structure of semi-natural and plantation stands of scots pine
  (pinus sylvestris l.) in northern scotland.
\newblock {\em Forestry}, 80:567--586.

\bibitem[Oliver and Larson, 1996]{oliver96}
Oliver, C.~D. and Larson, B.~C. (1996).
\newblock {\em Forest Stand Dynamics}.
\newblock John Wiley and Sons, New York, update edition.

\bibitem[Pacala et~al., 1996]{pacala96}
Pacala, S.~W., Canham, C.~D., Saponara, J., Silander, J.~A., Kobe, R.~K., and
  Ribbens, E. (1996).
\newblock Forest models defined by field measurements: Estimation, error
  analysis and dynamics.
\newblock {\em Ecological Monographs}, 66(1):1--43.

\bibitem[Penttinen et~al., 1992]{penttinen92}
Penttinen, A., Stoyan, D., and Henttonen, H.~M. (1992).
\newblock Marked point processes in forest statistics.
\newblock {\em Forest Science}, 38(4):806--824.

\bibitem[Perry et~al., 2003]{perry03}
Perry, L.~G., Neuhauser, C., and Galatowitsch, S.~M. (2003).
\newblock Founder control and coexistence in a simple model of asymmetric
  competition for light.
\newblock {\em Journal of Theoretical Biology}, 222:425--436.

\bibitem[Peterken, 1996]{peterken96}
Peterken, G.~F. (1996).
\newblock {\em Natural Woodland: Ecology and Conservation in Northern Temperate
  Regions}.
\newblock Cambridge University Press.

\bibitem[Raghib-Moreno, 2006]{raghib06}
Raghib-Moreno, M. (2006).
\newblock {\em Point Processes in Spatial Ecology}.
\newblock PhD thesis, University of Glasgow.

\bibitem[Renshaw et~al., 2009]{renshaw08}
Renshaw, E., Comas, C., and Mateu, J. (2009).
\newblock Analysis of forest thinning strategies through the development of
  space-time growth-interaction models.
\newblock {\em Stochastic Environmental Research and Risk Assessment},
  23(3):275--288.

\bibitem[Schneider et~al., 2006]{schneider06}
Schneider, M.~K., Law, R., and Illian, J.~B. (2006).
\newblock Quantification of neighbourhood-dependent plant growth by bayesian
  hierarchical modelling.
\newblock {\em Journal of Ecology}, 94:310--321.

\bibitem[Schutz, 2001]{schutz01}
Schutz, J. (2001).
\newblock Opportunities and strategies of transforming regular forests to
  irregular forests.
\newblock {\em Forest Ecology \& Management}, 151(1--3):87--94.

\bibitem[Schutz, 2002]{schutz02}
Schutz, J. (2002).
\newblock Silvicultural tools to develop irregular and diverse forest
  structures.
\newblock {\em Forestry}, 75(4):329--327.

\bibitem[Sinko and Streifer, 1967]{sinko67}
Sinko, J.~W. and Streifer, W. (1967).
\newblock A new model for age-size structure of a population.
\newblock {\em Ecology}, 48(6):910--918.

\bibitem[Stoll et~al., 1994]{stoll94}
Stoll, P., Weiner, J., and Schmid, B. (1994).
\newblock Growth variation in an naturally established population of pinus
  sylvestris.
\newblock {\em Ecology}, 75(3):660--670.

\bibitem[Taylor and MacLean, 2007]{taylor07}
Taylor, S.~L. and MacLean, D.~A. (2007).
\newblock Spatiotemporal patterns of mortality in declining balsam fir and
  spruce stands.
\newblock {\em Forest Ecology and Management}, 253:188--201.

\bibitem[Weiner et~al., 2001]{weiner01}
Weiner, J., Stoll, P., Muller-Landau, H., and Jasentuliyana, A. (2001).
\newblock The effects of density, spatial pattern, and competitive symmetry of
  size variation in simulated plant populations.
\newblock {\em The American Naturalist}, 158(4):438--450.

\bibitem[Wensel et~al., 1987]{wensel87}
Wensel, L., Meerschaert, W., and Biging, G. (1987).
\newblock Tree height and diameter growth models for northern california
  conifers.
\newblock {\em Hilgardia}, 55(8):1--20.

\bibitem[Wulder and Franklin, 2006]{wulder06}
Wulder, M.~A. and Franklin, S.~E. (2006).
\newblock {\em Understanding forest disturbance and spatial pattern: remote
  sensing and GIS approaches}.
\newblock CRC Press.

\bibitem[Wunder et~al., 2006]{wunder06}
Wunder, J., Bigler, C., Reineking, B., Fahse, L., and Bugmann, H. (2006).
\newblock Optimisation of tree mortality models based on growth partterns.
\newblock {\em Ecological Modelling}, 197:196--206.

\end{thebibliography}
\bibliographystyle{apalike}

\end{document}